\documentclass[twocolumn,prb,showpacs,preprintnumbers,superscriptaddress,amsmath,amssymb,citeautoscript,groupedaddress]{revtex4-1}

\pdfoutput=1

\usepackage{graphicx}
\usepackage{epstopdf}
\usepackage{dcolumn}
\usepackage{color}
\usepackage{amsmath,amssymb}
\usepackage[english]{babel}
\usepackage{bm}
\usepackage{multirow}
\usepackage[hyperindex]{hyperref}
\hypersetup{
	colorlinks,
	citecolor=blue,							
	filecolor=blue,							
	linkcolor=blue,
	urlcolor=blue}

\newcommand{\bpm}{\begin{pmatrix}}
\newcommand{\epm}{\end{pmatrix}}
\newcommand{\bs}{\boldsymbol}

\DeclareMathOperator{\rot}{rot}
\DeclareMathOperator{\diag}{diag}
\DeclareMathOperator{\tr}{tr}

\begin{document}

\title{Quantized Fermi-arc-mediated transport  in Weyl semimetal nanowires}
\author{Vardan Kaladzhyan}
\email{vardan.kaladzhyan@phystech.edu}
\author{Jens H. Bardarson}
\affiliation{Department of Physics, KTH Royal Institute of Technology, Stockholm, SE-106 91 Sweden}

\date{\today}

\begin{abstract}
We study longitudinal transport in Weyl semimetal nanowires, both in the absence and in the presence of a magnetic flux threading the nanowires.
We identify two qualitatively different regimes of transport with respect to the chemical potential in the nanowires.
In the ``surface regime", for low doping, most of the conductance occurs through the Fermi-arc surface states, and it rises in steps of one quantum of conductance as a function of the chemical potential; furthermore, with varying flux the conductance changes in steps of one quantum of conductance with characteristic Fabry-P\'erot interference oscillations. 
In the ``bulk-surface regime", for highly-doped samples, the dominant contribution to the conductance is quadratic in the chemical potential, and mostly conditioned by the bulk states; the flux dependence shows clearly that both the surface and the bulk states contribute.
The two aforementioned regimes prove that the contribution of Fermi arc surface states is salient and, therefore, crucial for understanding transport properties of finite-size Weyl semimetal systems.
Last but not least, we demonstrate that both regimes are robust to disorder.

\end{abstract}

\maketitle

%%%%%%%%%%%%%%%%%%%%%%%%%%%%%%%%%%%%%%%%%%%%%%%%%%%%%%%%%%%%%%%%%%
%%%%%%%%%%%%%%%%%%%%%%%%% Introduction %%%%%%%%%%%%%%%%%%%%%%%%%%%
%%%%%%%%%%%%%%%%%%%%%%%%%%%%%%%%%%%%%%%%%%%%%%%%%%%%%%%%%%%%%%%%%%

%
In 1929 the German mathematician and theoretical physicist Hermann Weyl proposed massless solutions of the Dirac equation \cite{Dirac1928}, the so-called ``Weyl fermions" \cite{Weyl1929}.
He demonstrated that in the absence of a mass term, the Dirac equation decoupled into two independent ones, also known as Weyl equations, each describing fermions of a given chirality, right or left.
Despite numerous theoretical predictions  \cite{Herring1937,Murakami2007,Wan2011,Burkov2011,Weng2015,Huang2015}, it was not until 2015 that Weyl fermions were first observed as the low-energy excitations in TaAs \cite{Xu2015a} and NbAs \cite{Xu2015b}.
Several decades ago Nielsen and Ninomiya showed that for continuous and periodic Hamiltonians with real spectra right and left chiralities are always bound to appear together \cite{Nielsen1981}.
Hence, a minimal low-energy model for Weyl fermions must embody both chiralities, and the numbers of particles with right and left chiralities must be equal.
The low-energy Weyl fermions of a given chirality $\lambda = \pm 1$ disperse linearly with momentum, $E^2_\lambda = v^2 \left[ p_x^2+p_y^2+(p_z - \lambda p_0)^2\right]$, with group velocity $v$ and band touching points $(0, 0, \pm p_0)$ referred to as ``Weyl points" or ``Weyl nodes", chosen without loss of generality to be along the $p_z$ axis.
In three dimensions the density of states of such quasiparticles grows quadratically with energy, while their group velocity is constant, and therefore, the bulk semiclassical conductance of Weyl semimetals $G \propto \mu^2$, where $\mu$ is the chemical potential of the sample, versus $G \propto \mu^{3/2}$ in ordinary metals.

Apart from this peculiar bulk property, Weyl semimetals are also known for their surface states --- Fermi arcs.
The contribution of these states to the transport properties of nanowires made of Weyl semimetals was considered both experimentally in Ref.~[\onlinecite{Nair2018}] and theoretically in Refs.~[\onlinecite{Baireuther2016,Gorbar2016,Baireuther2017,Igarashi2017,Fu2018,Deng2019,Breitkreiz2019}].
The experimental work mostly focuses on measuring the Shubnikov--de Haas effect, whereas theoretical papers contain semiclassical calculations of the conductance in different regimes.
\begin{figure}
	\centering
	\includegraphics[width=0.7\columnwidth]{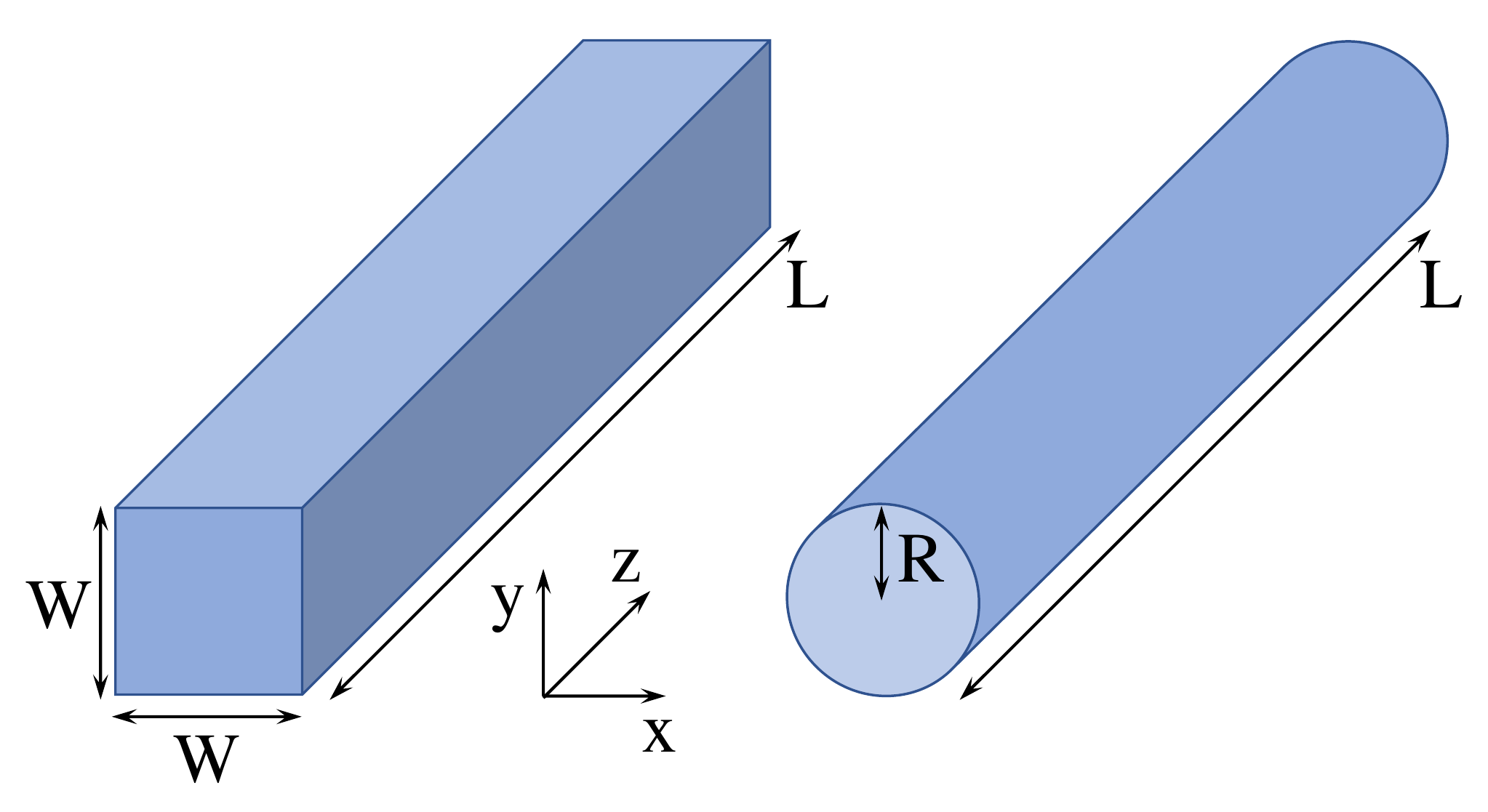}
	\caption{Weyl semimetal nanowires: (left) a slab of length $L$ with square cross-section $W\times W$; (right) a cylinder of length $L$ and radius $R$. For convenience, we use the slab for numerical simulations, whereas for analytical calculations we employ the cylinder.
	}
	\label{figsystem}
\end{figure} 

In this paper, we show that despite being three-dimensional \emph{per se}, Weyl semimetal nanowires may conduct only through the Fermi-arc surface states.
In order to demonstrate the latter, we calculate both analytically and numerically the zero-bias conductance of Weyl semimetal nanowires (see Fig.~\ref{figsystem}), taking into account the contribution of the Fermi-arc surface states.
We focus on studying transport properties of samples in which the transverse dimension of the system $W$ is, on one hand, much larger than the Fermi-arc localization length, but on the other hand, sufficiently small in order to resolve in energy the confinement gap appearing at the band touching points for the bulk states.
The conditions above ensure spatial separation of the bulk and the surface states, as well as experimental accessibility of the proposed regime \cite{Li2015,Wang2016a,Wang2016b,Wang2016c,Wang2018}.
We show that depending on the chemical potential in the sample there exist two qualitatively different regimes of conductance: ``surface regime" and ``bulk-surface regime".
In the latter the conductance of the nanowire is conditioned both by the bulk and by the surface states, and it grows quadratically with the chemical potential, showing the expected hallmark of three-dimensional linearly-dispersed electrons.
Surprisingly, in the former regime the nanowire shows effective one-dimensional behavior, and the conductance grows in steps of conductance quanta.
We explain such a remarkable feature by showing that in finite-size Weyl semimetal systems there is \emph{always} a window of energies, defined by the bulk confinement gap, where only surface states exist.
This inherent feature may serve as a strong evidence of the presence of the Fermi-arc surface states.
Furthermore, we study how the conductance is modified by magnetic flux penetrating the wire, and we demonstrate that in the surface regime it changes in steps of conductance quanta with characteristic Fabry-P\'erot interference oscillations, whereas in the bulk-surface regime the changes are not quantized.
Last but not least, we investigate the effects of weak and strong disorder on our results, and we find both regimes to be robust and our conclusions qualitatively unchanged.

%%%%%%%%%%%%%%%%%%%%%%%%%%%%%%%%%%%%%%%%%%%%%%%%%%%%%%%%%%%%%%%%%
%%%%%%%%%%%%%%%%%%%%%%%%%%%% Model %%%%%%%%%%%%%%%%%%%%%%%%%%%%%%
%%%%%%%%%%%%%%%%%%%%%%%%%%%%%%%%%%%%%%%%%%%%%%%%%%%%%%%%%%%%%%%%%

{\it Model.---} In what follows we perform transport calculations both numerically and analytically.  
For the former, we use the following cubic-lattice Hamiltonian adopted from Ref.~[\onlinecite{Behrends2019}]:
\begin{align}
\nonumber \mathcal{H}_{\mathrm{lat}} = v \left[\sin p_y\, \tilde{\sigma}_x - \sin p_x\, \tilde{\sigma}_y \right] \tilde{\tau}_z + v \sin p_z\,\tilde{\tau}_y + \phantom{aaa}\\
t \sum\limits_{i = x,y,z}(1-\cos p_i)\, \tilde{\tau}_x + v p_0\, \tilde{\sigma}_z,
\label{Hlatt}
\end{align}
where Pauli matrices $\bs{\tilde{\sigma}} = \{ \tilde{\sigma}_x,\,\tilde{\sigma}_y,\,\tilde{\sigma}_z\}$ and $\bs{\tilde{\tau}} = \{ \tilde{\tau}_x,\,\tilde{\tau}_y,\,\tilde{\tau}_z\}$ act in spin and orbital subspaces, respectively, $t$ denotes the hopping amplitude and $v$ parameterizes the low-energy velocity of Weyl fermions. 
We chose the $z$ direction to be the Weyl node separation axis with positions of the nodes given by $p_z = \pm p_0$.
For the sake of brevity we set $\hbar$ and the lattice constant $a$ to unity, restoring them in what follows if needed.
Note also that in all numerical simulations we chose $v=1$, $t=2/\sqrt{3}$.
This choice of parameters for the Hamiltonian in Eq.~(\ref{Hlatt}) yields two Weyl cones of velocity $v$ in the band structure, localized at $\pm p_0$.%

To perform analytical transport calculations we use a low-energy model with a block-diagonal form in the chirality subspace:
\begin{align}
\mathcal{H} = v p_z\, \sigma_z \tau_z - v p_0\, \sigma_z \tau_0 + v\left(p_x \sigma_x + p_y \sigma_y \right) \tau_z,
\label{Hanalyt}
\end{align}
where $\bs{\sigma} = \{ \sigma_x,\,\sigma_y,\,\sigma_z\}$ and $\bs{\tau} = \{\tau_x,\,\tau_y,\,\tau_z\}$ denote Pauli matrices acting in orbital and chirality subspaces, respectively. 
Since the Hamiltonian in Eq.~(\ref{Hanalyt}) is diagonal in the chirality subspace, for a given chirality $\lambda = \pm 1$ we can write a 2$\times$2 Hamiltonian as follows:
\begin{align}
\mathcal{H}_\lambda = \lambda v \left[ \left(p_z - \lambda p_0\right)\sigma_z + p_x \sigma_x + p_y \sigma_y \right].
\label{Hlambda}
\end{align}

In order to provide better understanding of transport properties of Weyl semimetal nanowires below we calculate the band structures of wires infinite in the $z$ direction with a finite cross-section in the $x$ and $y$ directions, in the absence of magnetic field.
For the lattice Hamiltonian in Eq.~(\ref{Hlatt}) it is sufficient to impose zero boundary conditions (also known as open or hard-wall boundary conditions), whereas for the Hamiltonian in Eq.~(\ref{Hanalyt}) it is necessary to derive boundary conditions, e.g., assuming a large-gap insulator outside of the wire \cite{Okugawa2014}. 
We leave the detailed derivation of boundary conditions to Appendix \ref{App:BCderivation}, presenting here the final result.
We consider a cylindrical wire of radius $R$ defined by $x^2+y^2 \leqslant R^2$ and we seek the solution of the Schr\"odinger equation for a given chirality $\mathcal{H}_\lambda \Psi_\lambda = E_\lambda \Psi_\lambda$ using the radial symmetry of the problem with the following ansatz
\begin{align}
\Psi_\lambda(r,\phi,z) = 
	\bpm
		\rho_+^\lambda(r) e^{i (m-1) \phi} \\
		\rho_-^\lambda(r) e^{i m \phi} 	
	\epm
	e^{i p_z z},
\end{align}
where $m \in \mathbb{Z}$ denotes the angular momentum quantum number, and $p_z$ is the good momentum in the $z$ direction. Radial functions $\rho^\lambda_{\pm}$ are defined as follows
\begin{align}
\rho^\lambda_-(r) &= J_m\left(\alpha r \right),\\
\rho^\lambda_+(r) &= \frac{i\lambda \alpha}{\lambda(p_z - \lambda p_0)-E_\lambda/v} J_{m-1}\left(\alpha  r \right),
\end{align}
where $\alpha \equiv \sqrt{E_\lambda^2/v^2 - (p_z - \lambda p_0)^2}$, $J_m(\dots)$ is the $m$-th Bessel function of the first kind.
The boundary condition thus reads
\begin{align}
\left[ \rho_-^\lambda(r) - i \lambda \rho_+^\lambda(r)\right] \Big|_{r = R} = 0.
\end{align}
The equation above yields the allowed energies for given values of $m$ and $p_z$, hence defining the band structure of an infinite Weyl nanowire.
Note that since chiralities are decoupled in Eq.~(\ref{Hanalyt}), the boundary condition does not mix chiralities either. 
\begin{figure}
	\centering
	\includegraphics[width=0.49\columnwidth]{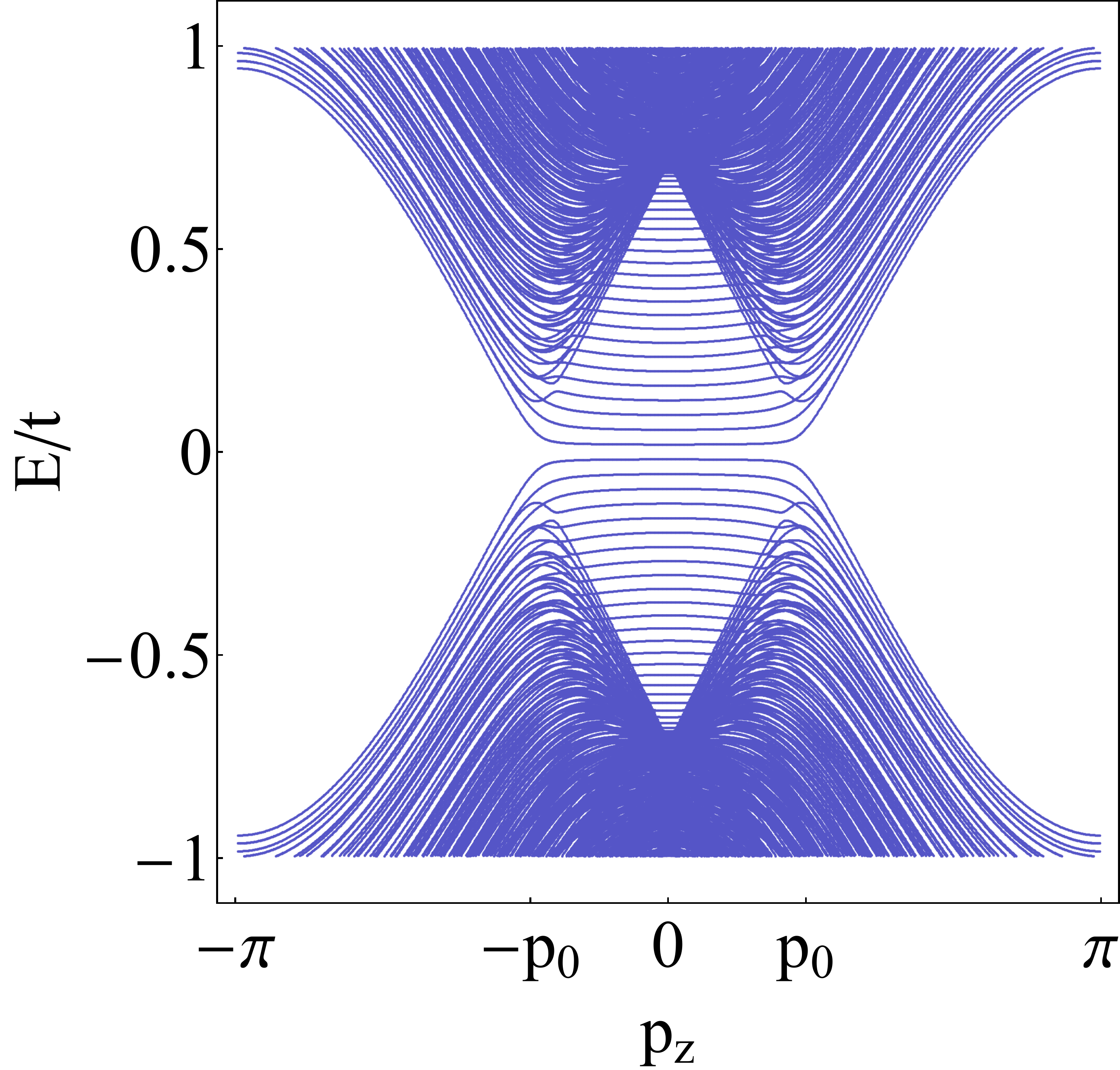}
	\includegraphics[width=0.49\columnwidth]{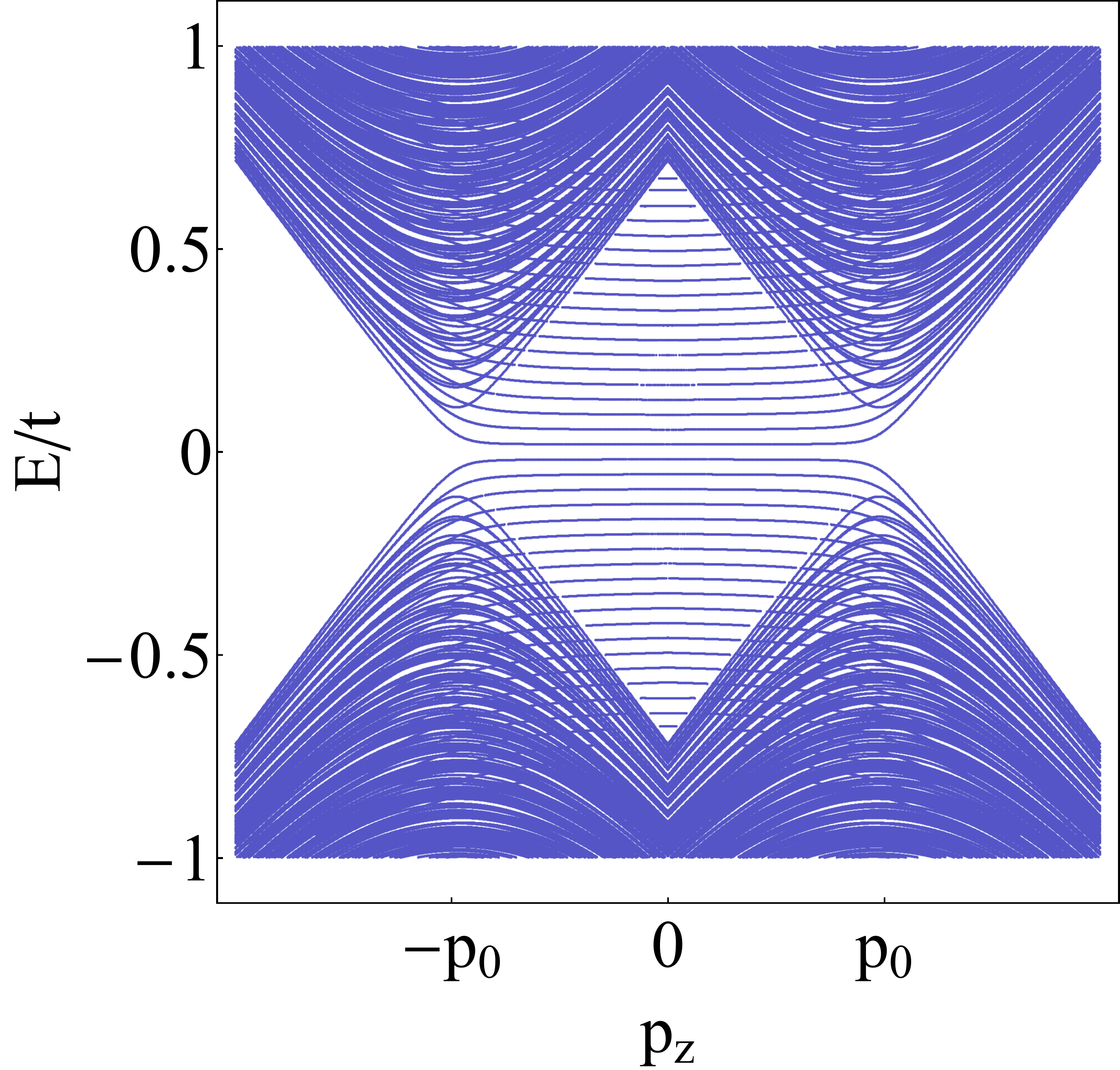}
	\caption{
Band structures computed numerically and analytically for a square slab of width $W=30$ (left) and for a cylindrical wire of radius $R=20$ (right), correspondingly (see Fig.~\ref{figsystem}).
We chose $p_0 = \pm 1$ for the positions of the nodes, and level spacing is given by $\Delta E \equiv 2\pi/4W \approx 2\pi/2\pi R = 0.05\,t$.
We assume that $t = 1\,$eV.
}
	\label{figbands}
\end{figure}

In Fig.~\ref{figbands} we plot band structures obtained numerically for a square cross-section slab and analytically for a cylinder (see Fig.~\ref{figsystem}). 
Nearly flat bands on both panels represent Fermi-arc surface states with characteristic localization length  $\ell_0 \equiv \hbar/2p_0$ \cite{Okugawa2014}.
It is worth noting that in the window of energies $\left| E \right| \lesssim 0.175\,t$ only surface solutions exist, whereas bulk solutions are gapped out due to finite-size effects.
The existence of such a window is conditioned by the relation between confinement gaps for the surface and the bulk states, namely, $\Delta_{\mathrm{surf}} < \Delta_{\mathrm{bulk}}$. 
Such a salient discrepancy stems from geometrical factors: indeed, the wave functions of the surface states are confined to a thin layer defined by the circumference of the cross-section of the wire, i.e., $4W$ for a slab and $ 2\pi R$ for a cylinder. 
The bulk states, however, have a smaller confinement length, namely, $W$ for a slab or $2R$ for a cylinder.
Hence the bulk confinement gap is larger than the surface one: $\Delta_{\mathrm{bulk}} = \hbar v/W$ or $\Delta_{\mathrm{bulk}} = \hbar v/2R$ versus $\Delta_{\mathrm{surf}} = \hbar v/4W$ or $\Delta_{\mathrm{surf}} = \hbar v/2\pi R$, correspondingly.
Therefore, we conclude that, remarkably, the window of energies with only surface solutions \emph{always} exists in finite-size Weyl semimetal systems.

%%%%%%%%%%%%%%%%%%%%%%%%%%%%%%%%%%%%%%%%%%%%%%%%%%%%%%%%%%%%%%%%%
%%%%%%%%%%%%%%%%%%%%%%%%% Conductance %%%%%%%%%%%%%%%%%%%%%%%%%%%
%%%%%%%%%%%%%%%%%%%%%%%%%%%%%%%%%%%%%%%%%%%%%%%%%%%%%%%%%%%%%%%%%

{\it Conductance.---}
Below we compute the zero-bias conductance of the nanowire both analytically, via the transfer matrix approach \cite{Tworzydlo2006,Bardarson2007,Xypakis2017}, and numerically, using the Kwant package \cite{Groth2014}.
Thus, we vary the chemical potential $\mu$ and the longitudinal magnetic field $B$, while calculating $G(\mu, B) = \lim_{V\to 0} I/V$, where $V$ is the bias, and $I$ is the current flowing through the wire. 
The analytical approach was described thoroughly in Refs.~[\onlinecite{Tworzydlo2006,Bardarson2007,Xypakis2017}], thus we leave the details of the calculation to Appendix \ref{App:conductance}.
We model the leads attached to the wire using the exact same Hamiltonians given in Eqs.~(\ref{Hlatt}) and (\ref{Hanalyt}), taken at the chemical potential $\mu_\infty$ large enough to emulate metallic electrodes.
In practice, it means that in the analytical low-energy model $\mu_\infty$ is taken to be larger than any other energy scale, whereas in the lattice model it must be smaller than the bandwidth, and should be chosen to yield the largest possible number of scattering states at $E=\mu_\infty$ in the lead.
Eventually $\mu_\infty$ drops out of all physically meaningful quantities such as, e.g., conductance.
We note also that analytically computed curves coincide with those obtained numerically, thus to avoid redundancy we restrict ourselves to presenting here only the numerical data, while leaving the analytical data to Appendix \ref{App:AnalyticalConductanceCurves}.

\begin{figure}
\begin{center}
	\includegraphics[width=\columnwidth]{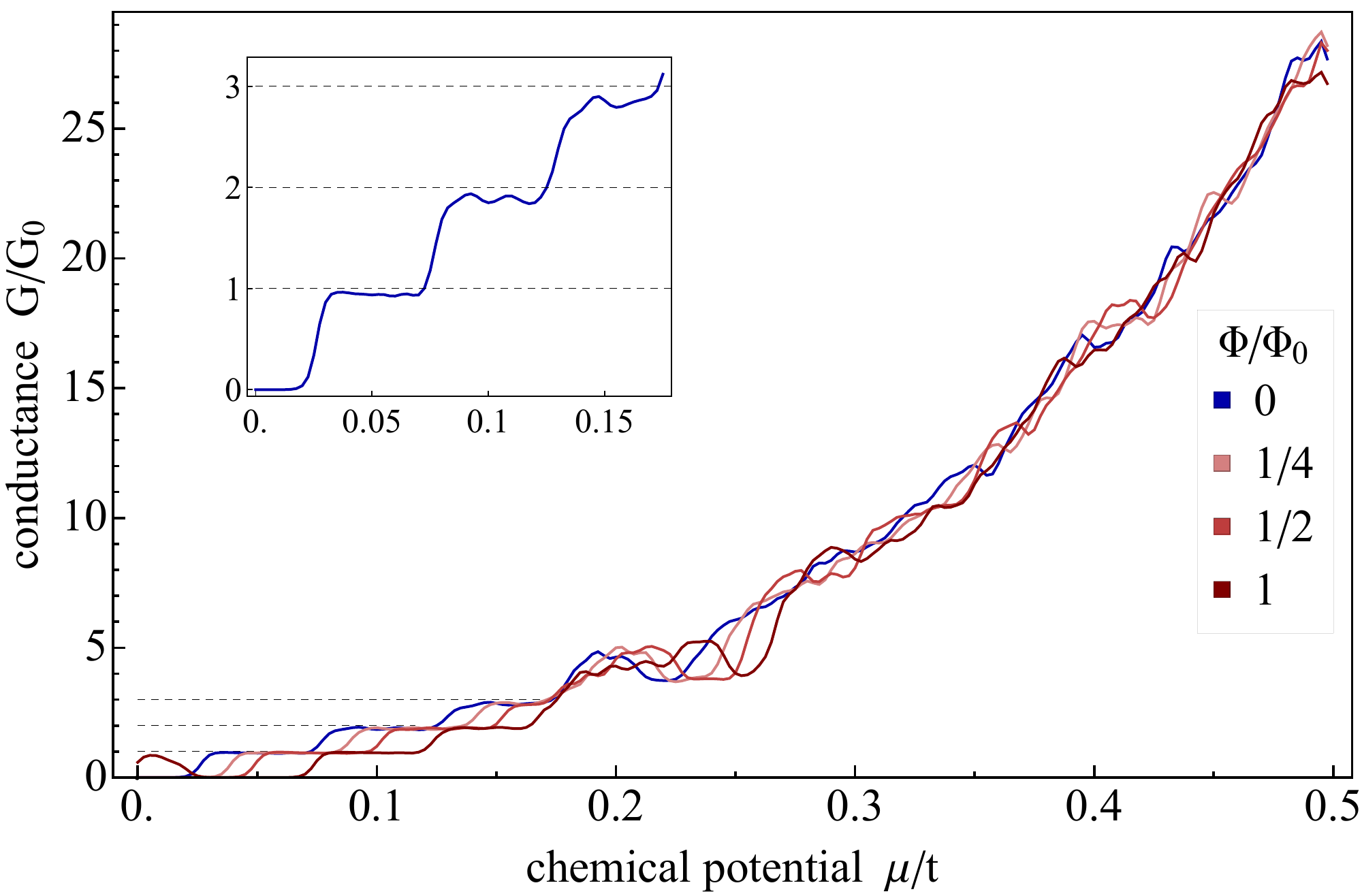}
	\caption{
Conductance of a Weyl semimetal slab $G$ in units of the conductance quantum $G_0$ plotted as a function of the chemical potential $\mu$ in units of the hopping amplitude $t$, for four different values of the magnetic flux $\Phi$.
Surface regime of conductance occurs at chemical potentials $\mu \lesssim 0.175\,t$, where the curve shows steps of $G_0$ characteristic of one-dimensional systems.
Contrary to that, for higher values of the chemical potential, we enter the bulk-surface regime with $\mu^2$ dependence characteristic of three-dimensional linearly-dispersed electrons.
Inset: conductance steps in the surface regime taken at $\Phi = 0$.
We can clearly see that the jumps on the curve take place at those values where the chemical potential crosses a new surface band (see Fig.~\ref{figbands}).
We added thermal broadening of $0.002\,t$ corresponding to $T \approx 23\,$K for $t = 1\,$eV.
} 
	\label{figmudepdifffluxes}
\end{center}
\end{figure}

In Fig.~\ref{figmudepdifffluxes} we plot the conductance of the wire in the units of the conductance quantum $G_0 \equiv e^2/h$ as a function of the chemical potential $\mu$ in the sample, for fixed values of the magnetic flux penetrating the wire, namely for $\Phi/\Phi_0 \in \{0,\, 1/4,\, 1/2,\, 1 \}$, where $\Phi_0 \equiv h/e$ is the quantum of flux.
We start by considering the case of zero flux, $\Phi = 0$.
It is worth noting that there are two qualitatively different regimes of conductance: ``surface regime" and ``bulk-surface regime".
In the surface regime transport occurs mainly through Fermi-arc surface states, with the characteristic feature being the conductance rising in steps of $G_0$ (see the inset in Fig.~\ref{figmudepdifffluxes}). 
This peculiar property can be elucidated as follows.
The dispersion of the surface states is effectively one-dimensional, i.e., their energy depends only on one of the two good momenta on the surface.
Therefore, the transport properties of the surface bands are similar to those of one-dimensional quantum wires (cf.\ Landauer formula \cite{Landauer1957}), and hence the conductance increases by $G_0$ every time the chemical potential crosses a new surface band.
The range of chemical potentials for the surface regime is defined by the finite-size gap, which in its turn is $ \propto \hbar v/R$, where $R$ is the radius of the wire \cite{Zhang2018}.
%
%We note also the remarkable agreement between analytical calculations and numerical simulations in Fig.~\ref{figmudepdifffluxes}.
%

%Above we have addressed the conductance of the nanowire in the absence of magnetic field.
%
Below we turn to the case of $\Phi \neq 0$.
In the absence of a Zeeman term, there are two main effects of the applied magnetic field on the band structure.
First, the orbital effects lead to the formation of Landau levels \cite{Abrikosov1998}, and second, all bands are shifted either down or up, depending on whether the magnetic field is parallel to the axis of the wire or antiparallel, respectively.
The reason for the latter is the fact that surface electrons have chiral dispersion, and therefore, their quasiclassical motion at the surface is clockwise or counterclockwise.
Since the magnetic field is applied perpendicular to the plane of this motion, it either favors their motion or not, depending on the direction of the field.
Thus, the more flux we apply the more we shift the band structure.	 

Several salient features of the conductance curves in Fig.~\ref{figmudepdifffluxes} for nonzero values of flux are worth being discussed.
First, it is clear that the shifts in the conductance curves for different values of flux are quantized in the surface regime (modulo interference oscillations) and irregular in the bulk regime.
Such difference stems from the fact that, as already mentioned earlier, in the surface regime most of the transport is conditioned by the Fermi arc states, which are localized in a thin layer of width $\ell_0$.
Such localization ensures that all surface electrons accumulate phases in a coherent manner.
Contrary to that, in the bulk-surface regime both the surface and the bulk states are responsible for transport. 
The transverse parts of the wave functions of the latter are localized at different distances from the axis of the wire, thus making distinct bulk states be affected by different values of the flux.
Such inhomogeneous influence of the magnetic field explains ``arbitrary" shifts of the  conductance curves for varying flux in the bulk-surface regime.

\begin{figure}[t!]
\begin{center}
	\includegraphics[width=0.48\columnwidth]{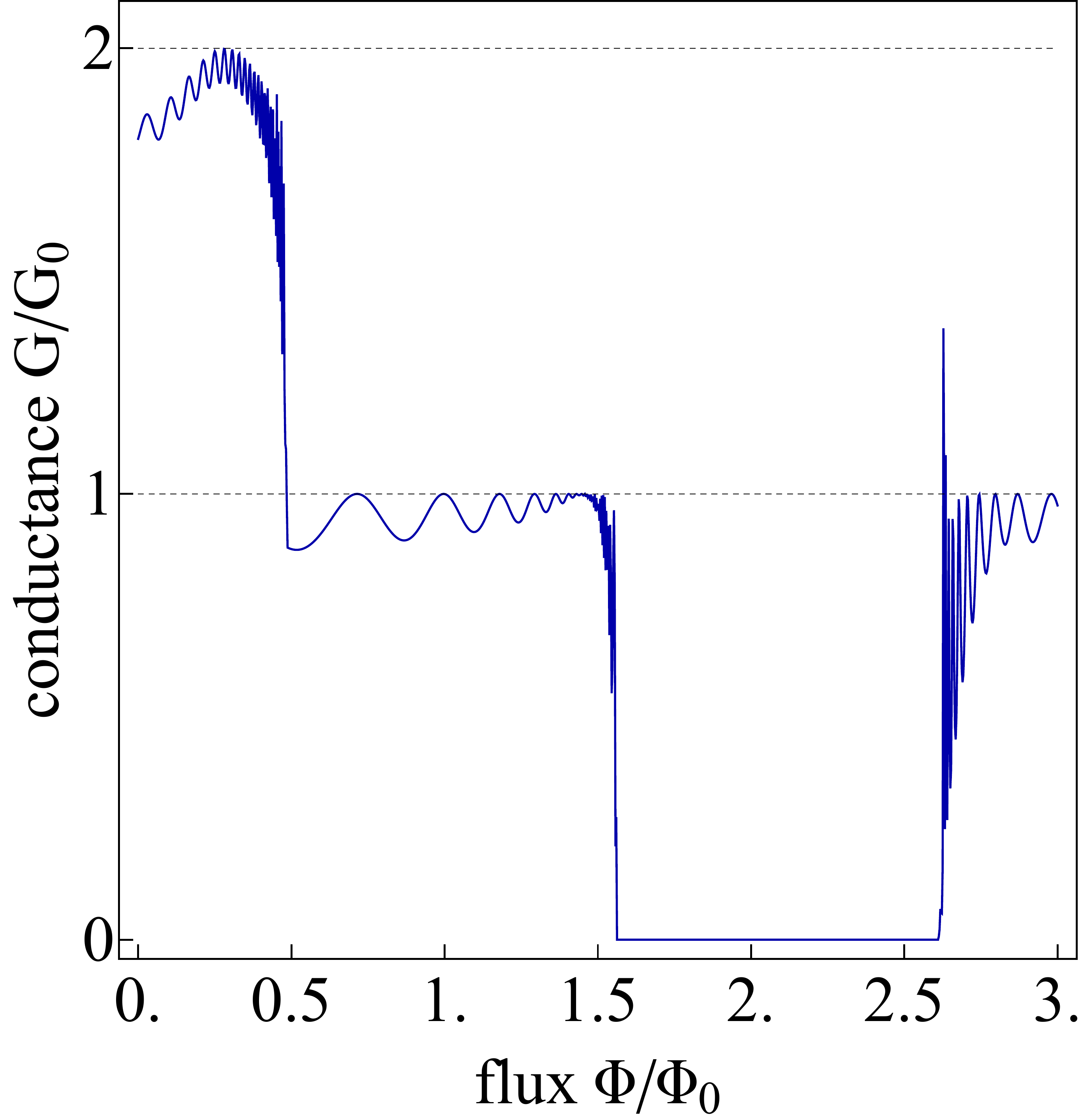}
	\includegraphics[width=0.496\columnwidth]{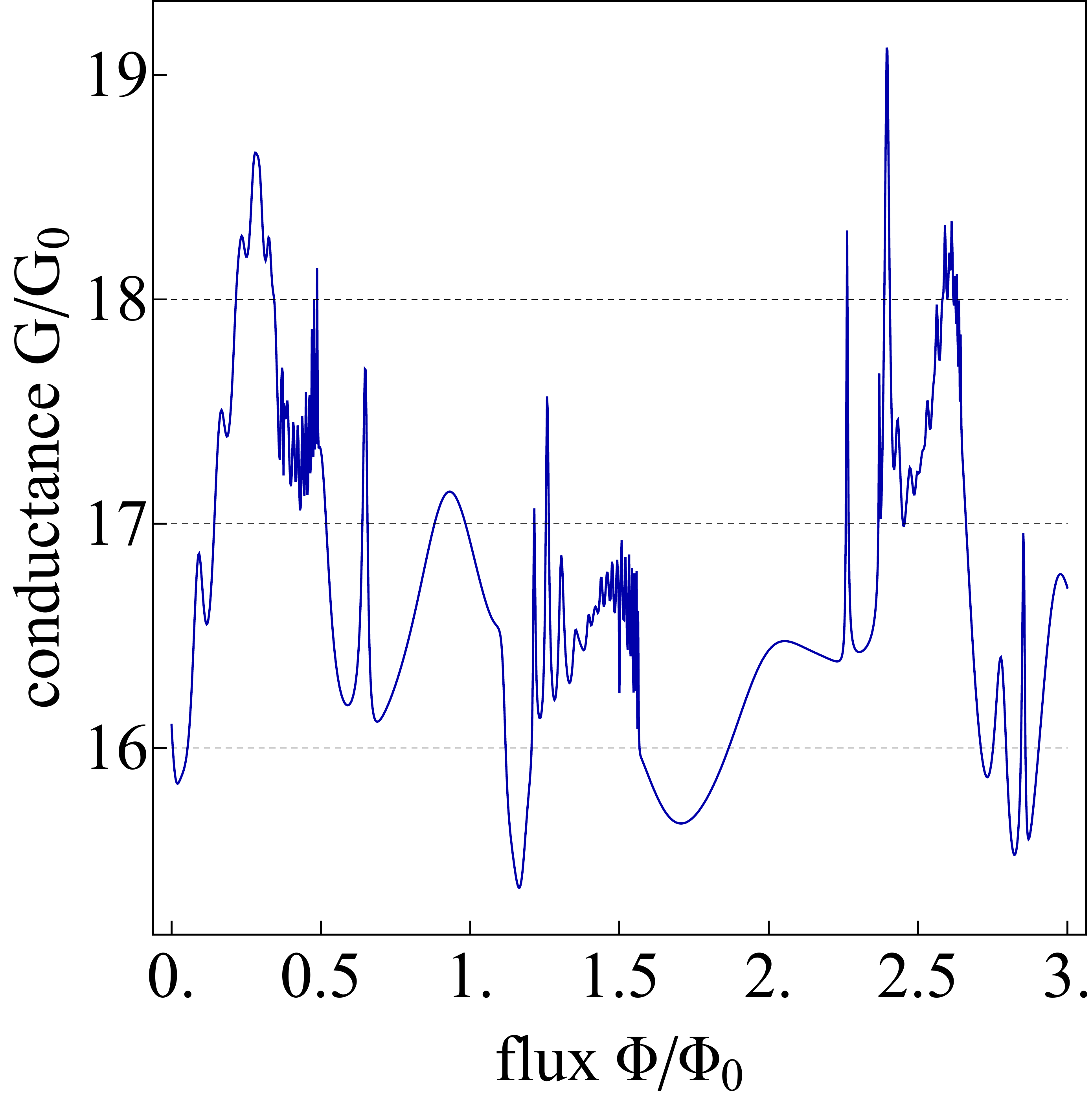}
	\caption{
Conductance of a Weyl semimetal slab $G$ in the units of conductance quantum $G_0$ plotted as a function of the magnetic flux $\Phi$ in the units of the flux quantum $\Phi_0$.
On the left and right panels we chose $\mu = 0.1\,t$ and $\mu = 0.4\,t$, corresponding to the surface and bulk-surface regimes, respectively (see Fig.~\ref{figmudepdifffluxes}).
In the surface regime conductance changes in steps of $G_0$ modulo Fabry-P\'erot interference oscillations, whereas in the bulk-surface regime the changes are irregular.
}
	\label{figfluxdep}
\end{center}
\end{figure}

Finally, it is both of theoretical and experimental interest to study how the conductance changes with the applied flux at fixed values of the chemical potential in the wire.
Previously, we have identified two qualitatively different regimes of transport depending on the chemical potential.
Thus, on the left and right panels in Fig.~\ref{figfluxdep} we plot the flux dependence of the conductance with the chemical potential fixed in the surface and bulk-surface regimes, correspondingly.

First, we analyze the surface regime.
As expected from the previous subsection, changes in conductance in that regime occur in steps of the conductance quantum modulo Fabry-P\'erot interference oscillations stemming from reflections from the leads.
The origin of these oscillations is easy to corroborate: it is sufficient to reduce the length of the wire by a factor of two, and verify that their period doubles; we have checked that this is indeed the case.
Second, we turn to the bulk-surface regime. Here, consistent with our antecedent findings, the conductance does not change in regular steps of $G_0$.
However, since the surface states still contribute to the transport, we can still identify the aforementioned interference oscillations.

\begin{figure}
	\centering
	\includegraphics[width=\columnwidth]{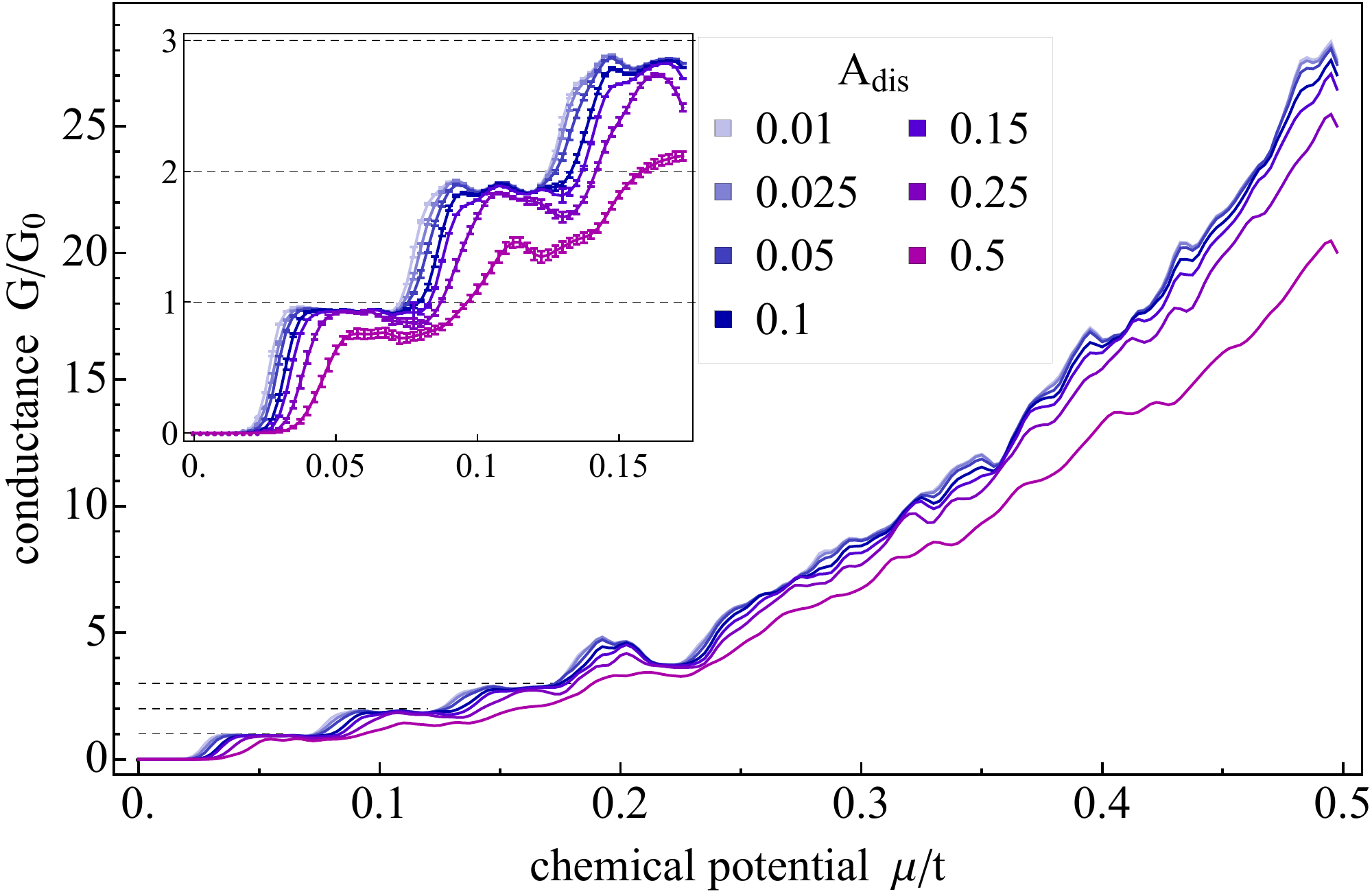}
	\caption{
Zero-flux conductance $G/G_0$ of a disordered Weyl semimetal slab as a function of the chemical potential $\mu$, averaged over 10 disorder realisations.
Error bars are intentionally omitted due to very small errors.
We take uniformly distributed delta-correlated disorder with maximum amplitude $A_{\mathrm{dis}}$ varying from $A_{\mathrm{dis}} = 0.01$ to $A_{\mathrm{dis}} = 0.5$.
It is clear that both regimes of conductance\textemdash surface and bulk-surface\textemdash survive up to high disorder strengths.
Inset: conductance steps in the surface regime averaged over 50 disorder realisations.
It is clear that for weak disorder, i.e., such that $ \frac{1}{3}\pi A_{\mathrm{dis}}^2 \lesssim \Delta E$, the quantized conductance steps are robust. 
All the curves are plotted with thermal broadening of $0.002\,t$ corresponding to $T \approx 23\,$K, assuming $t = 1\,$eV.
} 
	\label{figmudepdisorder}
\end{figure}

Last but not least, we have verified that our results hold in the presence of disorder by modeling the latter as a random uniform onsite variation of the chemical potential with amplitudes lying in $\left[ -A_{\mathrm{dis}},\, A_{\mathrm{dis}} \right]$.
In Fig.~\ref{figmudepdisorder} we present conductance curves for disordered samples with disorder amplitudes ranging from $A_{\mathrm{dis}} = 0.01$ to $A_{\mathrm{dis}} = 0.5$.
Average level broadening in the presence of such disorder can be estimated by $\Gamma \approx \frac{1}{3}\pi A^2_{\mathrm{dis}}$ \cite{Bruus2004}.
Thus, for $A_{\mathrm{dis}} = 0.25$ the level broadening $\Gamma \approx 0.065\,t$ becomes larger than the level spacing $\Delta E = 0.05\,t$.
Despite very strong disorder, both aforementioned regimes of conductance remain qualitatively unaffected.
The surface regime is robust due to the fact that the dispersion of the Fermi-arc surface states is effectively one-dimensional and chiral, and thus bereft of backscattering.
%

%%%%%%%%%%%%%%%%%%%%%%%%%%%%%%%%%%%%%%%%%%%%%%%%%%%%%%%%%%%%%%%%%%
%%%%%%%%%%%%%%%%%% Discussion and conclusions %%%%%%%%%%%%%%%%%%%%
%%%%%%%%%%%%%%%%%%%%%%%%%%%%%%%%%%%%%%%%%%%%%%%%%%%%%%%%%%%%%%%%%%

{\it Discussion and conclusions.---}
Above we have studied the longitudinal (magneto-)conductance of Weyl semimetal nanowires. 
First, we have found that depending on the chemical potential in the wire there exist two qualitatively different regimes of transport: surface regime and mixed bulk-surface regime.
In the former only the Fermi-arc surface states conduct, giving rise to quantized conductance steps, characteristic for one-dimensional physics.
Contrary to that, in the bulk-surface regime both the surface and the bulk states participate in transport, yielding the expected $G \propto \mu^2$ dependence.
%
%Due to such an exotic ambivalent behavior, Weyl semimetal nanowires can be thought of as bridges connecting one-dimensional and three-dimensional condensed matter physics. 
%\jhb{same comment about this sentence here as in main text}
%
Furthermore, we have investigated how the conductance varies with the magnetic flux penetrating the wire at fixed values of the chemical potential; we have shown that if the chemical potential is tuned to be in the surface regime, then there are regular jumps of one conductance quantum with characteristic interference oscillations, whereas in the mixed bulk-surface regime the conductance changes irregularly with the increasing value of the magnetic flux.

Despite being obtained for a particular model of a Weyl semimetal, our results can be generalized and applied to a broader range of models, both for Weyl and Dirac semimetals (see, e.g., transport experiments in Ref.~[\onlinecite{Nishihaya2019}]).
First, depending on the symmetries of a given material, realistic Weyl/Dirac semimetals may contain multiple pairs of Weyl cones at low energies, including spin-degenerate ones.
Qualitatively, this may lead to higher conductance values, however, further investigations with more realistic models are required to corroborate this hypothesis.
Moreover, for a Dirac semimetal with nonzero node separation in the presence of spin the height of the quantized conductance steps will double due to the spin degeneracy of the cones.
It is worth discussing also the case of time-reversal invariant Weyl semimetal nanowires.
In that case we have two pairs of cones in the Brillouin zone.
If both pairs of cones are separated in the $z$ direction, they generate two sets of Fermi-arc surface states related by time-reversal symmetry.
An applied magnetic flux shifts some of these states up in energy, whereas their time-reversal counterparts are shifted down.
Thus, as a function of the flux the confinement gap closes and reopens periodically, and therefore, in this case the analog of the left panel of Fig.~\ref{figfluxdep} will be periodic in flux, resembling the response of topological insulator nanowires \cite{Bardarson2010}.

Our results can be tested experimentally with relatively thin nanowires made of, e.g., TaAs, NbAs, TaP, NbP, GdPtBi, Co$_3$Sn$_2$S$_2$, etc., such that the confinement gap is sufficiently large to be resolved in energy \cite{Xu2015a,Xu2015b,Lv2015,Liu2015,Suzuki2016,Liu2018}.
Note also that realistic wires made of topological semimetals are generally larger than those considered in our work.
For instance, in Refs.~[\onlinecite{Li2015,Wang2016a,Wang2016b,Wang2016c,Wang2018}] the radii of the nanowires lie in the range of $30$ to $200$~nm, and the Fermi-arc localization length is of the order of $1$~nm.
To facilitate numerical simulations, our nanowires were taken to be approximately 10 times smaller than realistic ones, which means that in realistic systems the surface regime will occur in a smaller range of chemical potentials, namely $\mu \lesssim 10\,$meV.
While such small energies require higher resolution to be observed in larger samples, our theoretical conclusions will remain qualitatively unchanged.
Furthermore, in our geometry one quantum of flux through the wire is equivalent to having a magnetic field of $\approx 50\,$T.
In realistic wires one quantum of flux is achieved at smaller values of the magnetic field due to a larger cross-section area.

\begin{acknowledgments}
{\it Acknowledgments.} This work was supported by ERC Starting Grant No.~679722. V.K. would like also to thank Lo\"ic Herviou, David Aceituno, Jan Behrends, Emmanouil Xypakis, Sergue\"i Tchoumakov, Andrej Mesaros, Tobias Meng, Alexander Zyuzin and Mark O. Goerbig for fruitful discussions.
\end{acknowledgments}

\bibliography{biblio_WeylNanowires}

\widetext
\appendix

\section{Derivation of boundary conditions}\label{App:BCderivation}

In order to derive boundary conditions for a cylindrical Weyl nanowire we follow Okugawa and Murakami \cite{Okugawa2014}.
The approach is based on modeling the vacuum outside of the wire as an insulator with a gap $\Delta_\infty$ larger than all the other energy scales in the system, or formally, $\Delta_\infty \to -\infty$.
Thus, the system can be described by the following Hamiltonian
\begin{align}
\mathcal{H}_\lambda = 
\begin{cases}
	\bpm 
		\lambda (v p_z-\lambda \Delta)  & \lambda v (p_x - ip_y) \\
		\lambda v (p_x + i p_y) & -\lambda (v p_z-\lambda \Delta)
	\epm &\quad \text{for}\; r < R, \\
	\bpm 
		\lambda (v p_z-\lambda \Delta_\infty)  & \lambda v (p_x - ip_y) \\
		\lambda v (p_x + ip_y) & -\lambda (v p_z-\lambda \Delta_\infty)
	\epm &\quad \text{for}\; r > R, \\	
\end{cases}
\label{SM:HlambdaBC}
\end{align}
with $R$ being the nanowire radius, and $\lambda$ the given chirality.
We also denoted $\Delta \equiv v p_0$, where $p_0$ controls the positions of the Weyl nodes along the $p_z$ axis.
Rewriting $p_x \pm ip_y$ in polar coordinates as $-i \hbar e^{\pm i\phi} \left[\partial_r \pm \frac{i}{r}\partial_\phi \right]$ and using the radial symmetry of the problem, we employ an exponentially decaying ansatz for the wave function of a given chirality outside of the nanowire:
$$
\Psi_\lambda(r,\phi) \sim \bpm \rho^\lambda_+(r) e^{i(m-1)\phi} \\ \rho^\lambda_-(r) e^{i m \phi}\epm e^{-\varkappa_\lambda (r-R)},
$$
where $\varkappa_\lambda \equiv \sqrt{(v p_z-\lambda\Delta_\infty)^2 - E^2}/v > 0$ since $\Delta_\infty \to -\infty$. 
Inserting the ansatz above into the Schr\"odinger equation defined by the Hamiltonian in Eq.~(\ref{SM:HlambdaBC}), we get:
\begin{align*}
\lambda (v p_z-\lambda\Delta_\infty) \rho^\lambda_+ - i \lambda v \left[\partial_r + \frac{m}{r} - \sqrt{(v p_z-\lambda\Delta)^2 - E^2}/v \right] \rho^\lambda_- = E \rho^\lambda_+ \\
- i \lambda v \left[\partial_r - \frac{m-1}{r} - \sqrt{(v p_z-\lambda\Delta_\infty)^2 - E^2}/v \right] \rho^\lambda_+ - \lambda (v p_z-\lambda\Delta_\infty) \rho^\lambda_- = E \rho^\lambda_-
\end{align*}
We apply the limit of $\Delta_\infty \to -\infty$ to both equations above, and we end up with
\begin{align*}
\rho^\lambda_+ + i \lambda \rho^\lambda_- = 0 \\
i \lambda \rho^\lambda_+ - \rho^\lambda_- = 0.
\end{align*}
To have a non-trivial solution of the system above, we must satisfy $\rho^\lambda_+ + i\lambda \rho^\lambda_- = 0$, which in turn defines the boundary condition connecting radial parts of the wave function at $r=R$ in the limit of infinitely large $|\Delta_\infty|$:
\begin{align}
\rho^\lambda_-(R) - i \lambda \rho^\lambda_+(R) = 0
\label{SM:BC}
\end{align}
%It is worth noting that the same boundary condition can be derived by imposing that there is no current flowing through the boundary of the nanowire. 

\section{Weyl nanowire Band structure}
Below we consider a Weyl wire oriented along the $z$ axis. We use the Hamiltonian in Eq.~(2) and we introduce the magnetic field along the $z$ direction (i.e., along the line connecting the Weyl nodes) using the symmetric gauge: $\bs{\mathcal{A}} = B \left( -y,\,x,\, 0 \right)/2$, therefore $\bs{B} = \rot \mathcal{A} = B \hat{n}_z$.
Peierls substitution thus gives
\begin{equation} 
\mathcal{H} = 	
	\bpm 
		v (p_z - p_0) & v(\pi_x - i \pi_y) & 0 & 0 \\
		v(\pi_x + i \pi_y) & -v (p_z - p_0) & 0 & 0 \\
	0 & 0 & -v (p_z + p_0) & -v(\pi_x - i \pi_y) \\
	0 & 0 & -v(\pi_x + i \pi_y) & v (p_z + p_0) 
	\epm
\label{SM:Hwithmf}
\end{equation}
where $\pi_x = p_x - eBy/2, \pi_y = p_y + eBx/2$ with the electron charge given by $-e, e>0$.
It is also of use to rewrite this Hamiltonian in cylindrical coordinates using
$$
\pi_x \pm i \pi_y = -i \hbar e^{\pm i\phi} \left[\partial_r \pm \frac{i}{r}\partial_\phi \mp \frac{e B}{2\hbar}r \right], \quad \pi_x^2 + \pi_y^2 = -\hbar^2 \left[\partial_r^2 + \frac{1}{r}\partial_r + \frac{1}{r^2}\partial_\phi^2 \right] - i \hbar e B \partial_\phi +\frac{e^2B^2}{4} r^2
$$
as well as to compute the commutator
$$
\left[\pi_x, \pi_y \right] = \left[p_x - eBy/2, p_y + eBx/2 \right] = \frac{eB}{2} \left[p_x, x \right] - \frac{eB}{2} \left[y, p_y\right] = -i \hbar e B.
$$
To solve the Schr\"odinger equation we first note that problems for different chiralities are independent, therefore, we can rewrite it in a 2$\times$2 simplified form with chiralities $\lambda = \pm 1$:
\begin{equation}
\mathcal{H}_\lambda = \lambda v (p_z - \lambda p_0)\sigma_z  + \lambda v(\pi_x \sigma_x + \pi_y \sigma_y).
\end{equation} 
Equivalently in cylindrical coordinates: 
\begin{equation}
\mathcal{H}_\lambda = 
	\bpm
		\lambda v (p_z - \lambda p_0) &  -i \lambda \hbar v\,  e^{-i\phi} \left[\partial_r -\frac{i}{r}\partial_\phi + \frac{e B}{2\hbar}r \right] \\
		-i \lambda \hbar v\,  e^{+i\phi} \left[\partial_r +\frac{i}{r}\partial_\phi - \frac{e B}{2\hbar}r \right] & -\lambda v (p_z - \lambda p_0) 
	\epm
\end{equation}
We square the Hamiltonian and we get
\begin{equation*} 
	\bpm 
		v^2(p_z - \lambda p_0)^2 + v^2 (\pi_x^2 + \pi_y^2 + \hbar eB) & 0 \\
		0 & v^2(p_z - \lambda p_0)^2 + v^2 (\pi_x^2 + \pi_y^2 - \hbar eB)
	\epm \Psi_\lambda = E_\lambda^2 \Psi_\lambda, \quad \Psi_\lambda = \bpm \psi^\lambda_+ \\ \psi^\lambda_- \epm
\end{equation*}
For each component of the wave function we have:
\begin{equation}
\left[ -\hbar^2\left[\partial_r^2 + \frac{1}{r}\partial_r + \frac{1}{r^2}\partial_\phi^2 \right] - i \hbar e B \partial_\phi +\frac{e^2B^2}{4} r^2 + \sigma \hbar eB + (p_z - \lambda p_0)^2 - E^2/v^2 \right] \psi^\lambda_{\sigma} = 0.
\end{equation}
We use the following ansatz
$
\psi^\lambda_{\sigma} = \rho^\lambda_\sigma (r) e^{i (m - \Theta(\sigma) )\phi }  e^{i p_z z}
$
(where $\Theta$ is the Heaviside step function), and we get:
\begin{equation}
\left[ -\hbar^2\left[\partial_r^2 + \frac{1}{r}\partial_r - \frac{(m - \Theta(\sigma))^2}{r^2} \right] + \hbar e B (m - \Theta(\sigma)) +\frac{e^2B^2}{4} r^2 + \sigma \hbar eB + (p_z - \lambda p_0)^2 - E^2/v^2 \right] \rho^\lambda_\sigma(r) = 0.
\label{SM:eqtosolve}
\end{equation}
Below we solve Eq.~(\ref{SM:eqtosolve}) above in two different cases: $B = 0$ and $B \neq 0$. 

\subsection{Zero magnetic field}
At $B=0$ Eq.~(\ref{SM:eqtosolve}) simplifies to 
\begin{equation}
\left[ -\hbar^2\left[\partial_r^2 + \frac{1}{r}\partial_r  - \frac{(m - \Theta(\sigma))^2}{r^2} \right] + (p_z - \lambda p_0)^2 - E^2/v^2 \right] \rho^\lambda_\sigma(r) = 0.
\label{SM:eqnomf}
\end{equation}
We should also keep in mind that $\rho^\lambda_\sigma(r)$ are coupled via 
\begin{equation}
	\bpm 
		\lambda (p_z - \lambda p_0) & -i \lambda \hbar \left[\partial_r + \frac{m}{r} \right] \\
		-i \lambda \hbar \left[\partial_r - \frac{m-1}{r} \right] & -\lambda (p_z - \lambda p_0)
	\epm 
	\bpm
		\rho^\lambda_+(r) \\
		\rho^\lambda_-(r)
	\epm
	=
	\frac{E}{v}
	\bpm
		\rho^\lambda_+(r) \\
		\rho^\lambda_-(r)
	\epm.
\label{SM:eqnomfmatr}
\end{equation}
As long as $E^2/v^2 - (p_z - \lambda p_0)^2 \neq 0$, we can write a solution as
\begin{align}
\label{SM:rhominus}\rho^\lambda_-(r) &= J_m\left(\sqrt{E^2/v^2 - (p_z - \lambda p_0)^2}\; r /\hbar \right),\\
\label{SM:rhoplus}\rho^\lambda_+(r) &= \frac{i\lambda}{\lambda(p_z - \lambda p_0)-E/v} \sqrt{E^2/v^2 - (p_z - \lambda p_0)^2} J_{m-1}\left(\sqrt{E^2/v^2 - (p_z - \lambda p_0)^2}\; r /\hbar\right),
\end{align}
where $J_m$ stands for the $m$-th Bessel function of the first kind. The corresponding normalisation constant can be found with the help of the integral
$$
\int\limits_0^R r J^2_m\left(\alpha r \right) dr = \frac{R^2}{2} \left[ J^2_m\left(\alpha R \right) + J^2_{m+1}\left(\alpha R \right) \right] - \frac{m R}{\alpha} J_m\left(\alpha R \right) J_{m+1}\left(\alpha R \right).
$$

\subsection{Nonzero magnetic field}

In Eq.~(\ref{SM:eqtosolve}) we perform a change of variable as follows $\xi = e B r^2/2\hbar \equiv r^2/2\ell_B^2 $:
$$
\left\{ \xi \partial_\xi^2 + \partial_\xi + \left[-\frac{\xi}{4} + \frac{E^2/v^2-(p_z - \lambda p_0)^2-\sigma \hbar eB}{2\hbar eB} - \frac{m-\Theta(\sigma)}{2} - \frac{\left(m-\Theta(\sigma)\right)^2}{4\xi} \right] \right\}\tilde{\rho}^\lambda_\sigma(\xi) = 0.
$$
Below we set $\hbar = 1$ for the sake of simplicity. We use the two asymptotic limits of $\xi \to \infty$ and $\xi \to 0$ to build up a general solution, and we find
$$
\tilde{\rho}^\lambda_\sigma(\xi) = \tilde{C}_{m\sigma} e^{-\frac{\xi}{2}} \xi^{\frac{|m-\Theta(\sigma)|}{2}}\, _1F_1\left[ \frac{1+m-\Theta(\sigma) + |m-\Theta(\sigma)|}{2} - \frac{E^2/v^2-(p_z - \lambda p_0)^2-\sigma eB}{2eB},\,1+|m-\Theta(\sigma)|,\,\xi \right],
$$
where $_1F_1(a,b,z)$ is the confluent hypergeometric function of the first kind, and $\tilde{C}_{m\sigma}$ is a normalisation constant. Returning back to the original variable:
$$
\rho^\lambda_\sigma(r) = C_{m\sigma}\, e^{-\frac{eB r^2}{4}} r^{|m-\Theta(\sigma)|}\, _1F_1\left[ \frac{1+m-\Theta(\sigma) + |m-\Theta(\sigma)|}{2} - \frac{E^2/v^2-(p_z - \lambda p_0)^2-\sigma eB}{2eB},\,1+|m-\Theta(\sigma)|,\,\frac{eBr^2}{2} \right].
$$
It is worth noting that only one of the components of the wave function can be described by this expression. The other component should be found consistently using the original Schr\"odinger equation $\mathcal{H}_\lambda\Psi_\lambda = E \Psi_\lambda$. We choose here to express up to a normalisation constant the radial part of the lower component of the wave function:
\begin{equation}
\rho^\lambda_-(r) = C_{m-}e^{-\frac{eB r^2}{4}} r^{|m|}\, _1F_1\left[ \frac{1+m + |m|}{2} - \frac{E^2/v^2-(p_z - \lambda p_0)^2 + eB}{2 eB},\,1+|m|,\,\frac{eBr^2}{2} \right].
\label{Rminus}
\end{equation}
Next step is to find the upper component of the wave function using the lower one given by Eq.~(\ref{Rminus}). To do so we use the initial Schr\"odinger equation $\mathcal{H}_\lambda\Psi_\lambda = E \Psi_\lambda$, where we insert
\begin{equation}
\Psi_\lambda = 
	\bpm
		\rho^\lambda_+(r) e^{i (m-1)\phi} \\
		\rho^\lambda_-(r) e^{i m \phi}
	\epm
	e^{i p_z z}
\end{equation}

and we get
$$
	\bpm 
		\lambda (p_z - \lambda p_0) & -i\lambda \hbar  \left[\partial_r + \frac{m}{r} + \frac{e B}{2}r \right] \\
		-i \lambda \hbar  \left[\partial_r - \frac{m-1}{r} - \frac{e B}{2}r \right] & -\lambda (p_z - \lambda p_0)
	\epm 
	\bpm
		\rho^\lambda_+(r) \\
		\rho^\lambda_-(r)
	\epm
	=
	\frac{E}{v}
	\bpm
		\rho^\lambda_+(r) \\
		\rho^\lambda_-(r) 
	\epm.
$$
We use the equation above to express $\rho_+(r)$ in terms of found above $\rho_-(r)$:
\begin{align*}
\rho^\lambda_+(r) = \frac{i \lambda}{\lambda(p_z - \lambda p_0)-E/v}  \left[\partial_r + \frac{m}{r} + \frac{e B}{2}r \right]\rho_-(r), 
\end{align*}
for $E/v \neq \lambda(p_z - \lambda p_0)$. This very special case is never realized in finite-size samples. Thus, we get for $m>0$:
\begin{align*}
\nonumber \rho^\lambda_+(r) = \frac{2i \lambda}{\lambda(p_z - \lambda p_0)-E/v}  e^{-\frac{eB r^2}{4}} r^{m-1}\, \left\{m\, _1F_1\left[m - \frac{E^2/v^2-(p_z - \lambda p_0)^2}{2eB},\, 1+m,\, \frac{eBr^2}{2} \right] + \phantom{aaaaaaaaaaaa}\right. \\
\left. +\frac{eBr^2}{2} \frac{m - \frac{E^2/v^2-(p_z - \lambda p_0)^2}{2eB}}{1+m}\, _1F_1\left[1 + m - \frac{E^2/v^2-(p_z - \lambda p_0)^2}{2eB},\, 2+m,\, \frac{eBr^2}{2} \right] \right\}
\end{align*}
and 
\begin{align*}
\rho^\lambda_+(r) = \frac{i\lambda}{2} \left( \lambda(p_z - \lambda p_0) + E/v \right) \frac{1}{1-m} e^{-\frac{eB r^2}{4}} r^{1-m}\,\, _1F_1\left[1 - \frac{E^2/v^2-(p_z - \lambda p_0)^2}{2eB},\, 2-m,\, \frac{eBr^2}{2} \right]
\end{align*}
for $m \leqslant 0$. Using the boundary conditions in Eq.~(\ref{SM:BC}), we can calculate the bands in the presence of the magnetic field.

\section{Conductance of a cylindrical wire: transfer matrix approach}\label{App:conductance}

\subsection{Defining scattering states}

In order to compute the transfer matrix of the wire, we start by defining the scattering states of the problem.
We model the leads by means of the same Hamiltonian as the wire
\begin{align}
\mathcal{H}_\lambda = \lambda v (p_z - \lambda p_0)\sigma_z  + \lambda v(p_x \sigma_x + p_y \sigma_y),
\end{align} 
taken at a value of the chemical potential $\mu_\infty$ larger than any other energy scale of the system, and we normalise the scattering states in such a way that they carry unit current in the $z$-direction.
We note that regardless of whether we have or do not have a magnetic flux penetrating the wire, we set the magnetic field in the leads to zero, in order to simplify the calculation.
Thus, to compute the scattering momenta we solve the following equation:
\begin{align}
J_m\left(\sqrt{\mu_\infty^2/v^2 - (p_z - \lambda p_0)^2}\; R \right) + \frac{\sqrt{\mu_\infty^2/v^2 - (p_z - \lambda p_0)^2}}{\lambda(p_z - \lambda p_0)-\mu_\infty/v}  J_{m-1}\left(\sqrt{\mu_\infty^2/v^2 - (p_z - \lambda p_0)^2}\; R\right) = 0,
\label{SM:eqscatteringmomenta}
\end{align}
which defines a set of momenta $p_z = \left\{ p^i_\infty,\, i \in \overline{1,2N} \right\}$, to which we will henceforth refer to as ``scattering momenta".
We denoted the total number of scattering momenta as $2N$, because each chirality yields an equal number of scattering momenta.
Indeed, Eq.~(\ref{SM:eqscatteringmomenta}) is symmetric under changing $p_z \to - p_z$ and simultaneously changing $\lambda \to -\lambda$.
Knowing the scattering momenta, we can define a basis of scattering states:
\begin{align}
\Psi_{im}(r,\phi,z) = 
\begin{cases} 
	C_i^+	
	\bpm
		\rho_+^+(r) e^{i(m-1)\phi} \\
		\rho_-^+(r) e^{im\phi} \\
		0\\
		0
	\epm
	e^{i p^i_\infty z}\quad\text{for}\; p^i_\infty\;\text{coming from the positive chirality cone,} \\
	C_i^-		
	\bpm
		0\\
		0\\
		\rho_+^-(r) e^{i(m-1)\phi} \\
		\rho_-^-(r) e^{im\phi}
	\epm
	e^{i p^i_\infty z}\quad\text{for}\; p^i_\infty\;\text{coming from the negative chirality cone,} \\
\end{cases}
\label{SM:scatteringstates}
\end{align}
where $\rho_\pm^\lambda$ are defined in Eqs.~(\ref{SM:rhominus}) and (\ref{SM:rhoplus}).
In what follows, we order and normalize the scattering states in such a way that the first and the last $N$ states are right- and left-movers, respectively, yielding all unit currents, $+1$ and $-1$, correspondingly. The current operator is defined as $\hat{j_z} = v\, \sigma_z \otimes \tau_z$ (hereinafter for the sake of brevity we will omit the tensor product $\otimes$).
Chirality structure of the states in Eq.~(\ref{SM:scatteringstates}), as well as the angular parts and plane waves in the $z$-direction ensure that scattering states with different chiralities, scattering momenta and angular momenta are automatically orthogonal to each other.
Thus, to ensure that the states carry unit current for a given chirality we need to satisfy:
\begin{align}
\left| \underbrace{2\pi}_{\text{angles}} \times \underbrace{\lambda v}_{\text{current}} \times \left|C_i^\lambda \right|^2\int\limits_0^R rdr \left( \left|\rho_+^\lambda(r) \right|^2 - \left| \rho_-^\lambda(r)\right|^2 \right) \right| = 1,
\end{align}
which yields:
\begin{align}
\left|C_i^\lambda \right| = \left\{2\pi v \left|\frac{\alpha_i^2}{\gamma_i^2} I_{1}\left(m-1,\alpha_i,\alpha_i,R\right) - I_{1}\left(m,\alpha_i,\alpha_i,R\right) \right| \right\}^{-1/2},
\end{align}
where we defined $\alpha_i \equiv \sqrt{\mu_\infty^2/v^2 - (p^i_\infty - \lambda p_0)^2}$, $\gamma_i \equiv \lambda(p^i_\infty - \lambda p_0)-\mu_\infty/v$, and the function $I_1(m,\alpha,\beta,R)$ is defined in Appendix \ref{SM:AppIntegrals}.

Below we define the transverse part of the scattering states that will be consequently used to find the transfer matrix. Thus, uniting both chiralities to simplify notation, we define:
\begin{align}
\Phi_{im}(r,\phi) = 
	C_i^\lambda	
	\bpm
		\frac{1+\lambda}{2}\rho_+^+(r) e^{i(m-1)\phi} \\
		\frac{1+\lambda}{2}\rho_-^+(r) e^{im\phi} \\
		\frac{1-\lambda}{2}\rho_+^-(r) e^{i(m-1)\phi} \\
		\frac{1-\lambda}{2}\rho_-^-(r) e^{im\phi}
	\epm
	\quad
	\text{or}
	\quad
\Phi^\lambda_{im}(r,\phi) = 
	C_i^\lambda	
	\bpm
		\rho_+^\lambda(r) e^{i(m-1)\phi} \\
		\rho_-^\lambda(r) e^{im\phi} 
	\epm.
\label{SM:scatteringstatestransverse}
\end{align}

\subsection{Transfer matrix approach}
We define a basis of scattering states 
$$
\{\Psi_{1m},\Psi_{2m} \dots \Psi_{N\,m}, \Psi_{N+1\,m} \Psi_{N+2\,m} \dots \Psi_{2N\,m} \} 
$$
and order the states in such a way that the first $N$ states are right-movers carrying unit currents from left to right, and the last $N$ states are left-movers carrying unit currents from right to left.
We can write this formally in terms of the transverse parts as
\begin{align}
\int\limits_0^R rdr \int\limits_0^{2\pi}d\phi\; \Phi_{im}^\dag(r,\phi) \cdot v\, \sigma_z \tau_z \cdot \Phi_{jm'}(r,\phi) = D_{ij}\delta_{mm'}, \quad\text{where}\; i,j \in \overline{1,2N}\;\text{and}\; m,m' \in \mathbb{Z}.
\label{SM:normalisationforPhi}
\end{align}
In the equation above $D_{ij}$ is a diagonal $2N\times 2N$ matrix defined as follows:
$$
D \equiv || D_{ij} || = \diag ( \underbrace{+1\;+1\;\dots +1}_{N}\; \underbrace{-1\;-1\;\dots -1}_{N}).
$$

We introduce the magnetic field parallel to the $z$ direction into the sample and we write down the Schr\"odinger equation for the wave function in the sample:
$$
\left[ v p_z\, \sigma_z \tau_z - \Delta \sigma_z \tau_0 + v\left(\pi_x \sigma_x + \pi_y \sigma_y \right) \tau_z - \mu \sigma_0 \tau_0 \right] \Psi(r,\phi,z) = 0.
$$
We can represent the wave function in the basis of the scattering states $\Phi_{im}(r,\phi)$ defined above:
$$
\Psi(r,\phi,z) = \sum\limits_{jm'} \zeta_{jm'}(z) \Phi_{jm'}(r,\phi),
\quad \text{where} \quad 
\zeta_{jm'}(z) \equiv \int\limits_0^R rdr \int\limits_0^{2\pi}d\phi\; \Phi^\dag_{jm'}(r,\phi)\Psi(r,\phi,z). 
$$
We multiply our equation by $\Phi^\dag_{im}(r,\phi)$ from the left, and we integrate it over all radii and angles:
$$
\int\limits_0^R rdr \int\limits_0^{2\pi}d\phi\;\sum\limits_{jm'} \Phi^\dag_{im}(r,\phi) \Big[ v p_z\, \sigma_z \tau_z - \Delta \sigma_z \tau_0 + v\left(\pi_x \sigma_x + \pi_y \sigma_y \right) \tau_z - \mu \sigma_0 \tau_0 \Big] \Phi_{jm'}(r,\phi) \zeta_{jm'}(z)  = 0.
$$
Thus we get:
$$
\sum\limits_{jm'} D_{ij} \delta^{mm'}p_z \zeta_{jm'}(z) + \delta^{mm'} U^m_{ij} \zeta_{jm'}(z) = 0 \quad\Rightarrow\quad \sum\limits_{j} D_{ij} p_z \zeta_{jm}(z) +  U^m_{ij} \zeta_{jm}(z) = 0,
$$
where we denoted
$$
U^m_{ij} = \int\limits_0^R rdr \int\limits_0^{2\pi}d\phi\;\Phi^\dag_{im}(r,\phi) \Big[ - \Delta \sigma_z \tau_0 + v\left(\pi_x \sigma_x + \pi_y \sigma_y \right) \tau_z - \mu \sigma_0 \tau_0 \Big] \Phi_{jm}(r,\phi).
$$
The exact analytical expression for the matrix $U^m \equiv ||U_{ij}^m||$ is calculated in Appendix \ref{SM:AppMatrixElementU}.
The equations above show that there is no term mixing between different states with $m$ and $m'$ while they propagate through the sample, therefore, we can compute the transfer matrix for channel $m$, and then sum up corresponding conductances over all relevant values of $m$.
In order to simplify further calculations, we introduce a vector 
$$
\vec{\zeta}_m(z) \equiv (+\zeta_{1m}(z) \dots +\zeta_{N\,m}(z)\;-\zeta_{N+1\,m}(z)\dots -\zeta_{2N\, m}(z)\, )^\mathrm{T} 
$$ 
and, replacing $p_z \to -i \hbar \partial_z$, rewrite the equation above as
$$
\partial_z \vec{\zeta}_m(z) + \frac{i}{\hbar} \tilde{U}^m \vec{\zeta}_m(z) = 0,
$$
where $\tilde{U}^m \equiv U^m \cdot D$.
Assuming that $L$ is the length of the scattering region, we find the transfer matrix as matrix exponential in the following form:
\begin{align}
T_m = \exp \left\{ -\frac{i}{\hbar} \tilde{U}^m L \right\}.
\label{SM:Tmatrix}
\end{align}
We must ensure that this expression is current-conserving, i.e., $T^\dag_m \cdot D \cdot T_m = D$. 
The latter is easy to verify using the definition of the matrix exponential and mathematical induction, and we leave this proof to Appendix \ref{SM:AppCurrentConservationProof}.

Finally, to find the conductance we use the Landauer formula.
The transfer matrix for a given channel $m$ is a $2N\times 2N$ matrix of the following form \cite{Mello1988}
\begin{align}
T_m = 
	\bpm
		\left(t_m^\dag \right)^{-1} & r_m' \left( t_m'\right)^{-1} \\
		- \left( t_m'\right)^{-1} r_m & \left( t_m'\right)^{-1}
	\epm,
\end{align}
where $t_m$ and $t_m'$ are the transmission matrices from left to right and right to left, respectively, and $r_m$ and $r_m'$ are the corresponding reflection matrices. 
Thus, the transmission matrix $t_m$ is an $N\times N$ matrix, and can be defined as follows using the transfer matrix in Eq.~(\ref{SM:Tmatrix}):
\begin{align}
\left(t^\dag_m\right)^{-1} = 
\bpm \left(T_m\right)_{11} & \dots & \left(T_m\right)_{1\,N} \\
\dots & \dots & \dots \\
 \left(T_m\right)_{N\,1} & \dots & \left(T_m\right)_{N\,N} \epm
\end{align}
The conductance of the corresponding channel $m$ and the full conductance are, therefore, given by
\begin{align}
G_m = G_0 \tr \left( t^\dag_m t_m \right), \quad G = \sum\limits_{m \in \mathbb{Z}} G_m,
\end{align}
where $G_0 \equiv e^2/h$ is the conductance quantum.

\section{Integrals for T-matrix calculation}\label{SM:AppIntegrals}

\noindent In order to compute the overlap integrals we need to compute only the two following integrals:
\begin{enumerate}
\item $I_1(m,\alpha,\beta,R) = \int_0^R r J_m\left(\alpha r \right) J_m\left(\beta r \right) dr$,
\item $I_2(m,\alpha,\beta,R) = \int_0^R r^2 J_m\left(\alpha r \right) J_{m-1}\left(\beta r \right) dr$,
\end{enumerate}
where $\alpha, \beta \neq 0$. We start with the first integral that has a closed analytical form. Integrating by parts we get:
\begin{align*}
\int_0^R r J_m\left(\alpha r \right) J_m\left(\beta r \right) dr = 
\begin{cases}
	\frac{R}{\alpha^2-\beta^2} \left[\beta J_m\left(\alpha R \right) J_{m-1}\left(\beta R \right) - \alpha J_m\left(\beta R \right) J_{m-1}\left(\alpha R \right) \right] \quad &\text{for}\quad \alpha^2 \neq \beta^2 \\
	(-1)^{2^{\delta_{0,\alpha-\beta}}m}\left\{\frac{R^2}{2} \left[ J^2_m\left(\alpha R \right) + J^2_{m+1}\left(\alpha R \right)\right]	 - \frac{m R}{\alpha} J_m\left(\alpha R \right) J_{m+1}\left(\alpha R \right) \right\} \quad &\text{for}\quad \alpha^2 = \beta^2 %\neq 0 \\
%	\frac{R^2}{2} \delta_{0,m} \quad &\text{for}\quad \alpha = \beta = 0 
	\end{cases}
\end{align*}
where $\delta_{ij}$ stands for the Kronecker's delta symbol. The second integral does not have an analytical form for the most general case. So below we first restrict ourselves to cases where explicit analytical expressions can be derived. We start with the case where $\alpha^2 = \beta^2$:
\begin{align*}
	\int_0^R r^2 J_m\left(\alpha r \right) J_{m-1}\left(\beta r \right) dr = 
	(-1)^{2^{\delta_{0,\alpha-\beta}}(m-1)} \frac{R}{2\alpha^2} \times \phantom{aaaaaaaaaaaaaaaaaaaaaaaaaaaaaaaaaaaaaaaaaaaaaaaaaaaaaa} \\
	\times
	\begin{cases}	
	\alpha R \left[  m J^2_{m}\left(\alpha r \right) + (m-1)J^2_{m+1}\left(\alpha r \right) \right] -2m(m-1)J_{m}\left(\alpha r \right)J_{m+1}\left(\alpha r \right) \quad &\text{for}\quad m \geqslant 0  \\
	\alpha R \left[ m J^2_{m-2}\left(\alpha r \right) + (m-1) J^2_{m-1}\left(\alpha r \right) \right] -2m(m-1)J_{m-2}\left(\alpha r \right)J_{m-1}\left(\alpha r \right) \quad &\text{for}\quad m < 0
	\end{cases}
\end{align*}
%If $\alpha = \beta = 0$ then we have
%\begin{align*}
%	\int_0^R r^2 J_m\left(\alpha r \right) J_{m-1}\left(\beta r \right) dr = 
%	\frac{R^3}{3} \delta_{0,m}\, \delta_{0, m-1} = 0. \phantom{aaaaaaaaaaaaaaaaaaaaaaaaaaaaaaaaaaaaaaaaaaaaaaaaaaaaaaaaa}
%\end{align*}
Finally, in the case of $\alpha^2 \neq \beta^2$ we can write down a recursive relation expressing the integral at arbitrary $m$ in terms of two integrals at $m=1$ and $m=0$ that can be calculated exactly due to symmetries and properties of Bessel functions:
\begin{align*}
I_2(0,\alpha,\beta,R) &= \frac{\alpha}{\beta} I_2(1,\alpha,\beta,R) + \frac{R^2}{\beta} J_0(\alpha R) J_0(\beta R) - \frac{2}{\beta} I_1(0,\alpha,\beta,R) \\
I_2(1,\alpha,\beta,R) &= -\frac{1}{(\alpha^2-\beta^2)^2} \Bigg\{ \alpha R J_0(\beta R) \left[(\alpha^2-\beta^2)R J_0(\alpha R) - 2\alpha J_1(\alpha R) \right] + \beta R J_1(\beta R) \left[(\alpha^2-\beta^2)R J_1(\alpha R) + 2\alpha J_0(\alpha R) \right] \Bigg\} 
\end{align*}
And finally, the recursive equation for $I_2(m,\alpha,\beta,R)$:
\begin{align*}
\nonumber I_2&(m,\alpha,\beta,R) = \\
	&=\begin{cases}
		\frac{2\beta}{\alpha} \left[I_2(m-1,\alpha,\beta,R) + \frac{\alpha}{2\beta} I_2(m-1,\beta,\alpha,R) - \frac{R^2}{\beta} J_{m-1}(\alpha R) J_{m-1}(\beta R) - \frac{m-3}{\beta}I_1(m-1,\alpha,\beta,R) \right], &m \geqslant 2 \\
		\frac{2\alpha}{\beta} \left[I_2(m+1,\alpha,\beta,R) + \frac{\beta}{2\alpha} I_2(m+1,\beta,\alpha,R) + \frac{R^2}{\alpha} J_{m}(\alpha R) J_{m}(\beta R) - \frac{m+2}{\alpha}I_1(m,\alpha,\beta,R) \right], &m \leqslant 0
	\end{cases}
\end{align*}

\section{Calculating matrix elements \boldmath{$U_{ij}^m$}}\label{SM:AppMatrixElementU}
Below we calculate the matrix elements of the Hamiltonian in the scattering states:
\begin{align}
U^m_{ij} = \int\limits_0^R rdr \int\limits_0^{2\pi}d\phi\;\Phi^\dag_{im}(r,\phi) \Big[ - \Delta \sigma_z \tau_0 + v\left(\pi_x \sigma_x + \pi_y \sigma_y \right) \tau_z - \mu \sigma_0 \tau_0 \Big] \Phi_{jm}(r,\phi).
\end{align}
Since the Hamiltonian is diagonal in the chirality subspace (i.e., if we chose $\Phi^\dag_{im}(r,\phi)$ and $ \Phi_{jm}(r,\phi)$ to be of different chiralities, then we would get zero), we can simplify it to:
\begin{align}
U^m_{ij} = \int\limits_0^R rdr \int\limits_0^{2\pi}d\phi\;\left[\Phi^\lambda_{im}(r,\phi)\right]^\dag \Big[ - \mu \sigma_0 - \Delta \sigma_z  + \lambda v\left(\pi_x \sigma_x + \pi_y \sigma_y \right)  \Big] \Phi^\lambda_{jm}(r,\phi).
\end{align}
The first two terms are constants, and therefore, we get:
\begin{align}
\nonumber\int\limits_0^R rdr \int\limits_0^{2\pi}d\phi\;\left[\Phi^\lambda_{im}(r,\phi)\right]^\dag \Big[ - \mu \sigma_0 - \Delta \sigma_z  \Big] \Phi^\lambda_{jm}(r,\phi) = \phantom{aaaaaaaaaaaaaaaaaaaaaaaaaaaaaaaaaaaaaaaa}\\
= 2\pi C_i^\lambda C_j^\lambda \left[(-\mu-\Delta) \frac{\alpha_i \alpha_j}{\gamma_i \gamma_j} I_1\left(m-1, \alpha_i, \alpha_j, R \right) + (-\mu+\Delta) I_1\left(m, \alpha_i, \alpha_j, R \right) \right],
\label{SM:muanddelta}
\end{align}
where $I_1\left(m,\alpha,\beta,R \right)$ is defined in Appendix \ref{SM:AppIntegrals}, and $\alpha_i, \gamma_i$ are given in the subsection where we calculated the scattering states. We deal with the last term by rewriting it in polar coordinates:
\begin{align}
\nonumber &-i\lambda \hbar v \int\limits_0^R rdr \int\limits_0^{2\pi}d\phi\;\left[\Phi^\lambda_{im}(r,\phi)\right]^\dag \bpm 0 & e^{-i\phi} \left(\partial_r - \frac{i}{r}\partial_\phi + \frac{eB}{2\hbar}r \right)\\ e^{+i\phi} \left(\partial_r + \frac{i}{r}\partial_\phi - \frac{eB}{2\hbar}r \right) & 0 \epm \Phi^\lambda_{jm}(r,\phi) = \phantom{aaaaaaaaaaaaaaaaaaaaaaa}\\
\nonumber & -2\pi i\lambda \hbar v C_i^\lambda C_j^\lambda \int\limits_0^R \negthickspace rdr \left\{ \left[ \rho_+^\lambda(p_\infty^i, r) \right]^* \left(\partial_r + \frac{m}{r} + \frac{eBr}{2\hbar} \right) \rho_-^\lambda(p_\infty^j, r) + \left[ \rho_-^\lambda(p_\infty^i, r) \right]^* \left(\partial_r - \frac{m-1}{r} - \frac{eBr}{2\hbar} \right) \rho_+^\lambda(p_\infty^j, r) \right\} = \\
\nonumber & -2\pi \hbar v C_i^\lambda C_j^\lambda \int\limits_0^R \negthickspace rdr \left\{ \alpha_j \left[ \frac{\alpha_i}{\gamma_i}J_{m-1}(\alpha_i r)J_{m-1}(\alpha_j r) + \frac{\alpha_j}{\gamma_j}J_{m}(\alpha_i r)J_{m}(\alpha_j r)\right] + \frac{eBr}{2\hbar} \left[  \frac{\alpha_i}{\gamma_i}J_{m}(\alpha_j r)J_{m-1}(\alpha_i r) + \mathrm{id.}\, i \leftrightarrow j\right] \right\} = \\
& -2\pi \hbar v C_i^\lambda C_j^\lambda \left\{\negthickspace\alpha_j \negthickspace\left[ \frac{\alpha_i}{\gamma_i}I_1\left(m-1,\alpha_i,\alpha_j, R \right)\negthickspace +\negthickspace \frac{\alpha_j}{\gamma_j}I_1\left(m,\alpha_i,\alpha_j, R \right)\right]\negthickspace +\negthickspace \frac{eB}{2\hbar} \negthickspace \left[ \frac{\alpha_i}{\gamma_i}I_2\left(m,\alpha_j,\alpha_i, R \right)\negthickspace +\negthickspace \frac{\alpha_j}{\gamma_j}I_2\left(m,\alpha_i,\alpha_j, R \right)\right] \negthickspace\right\}
\label{SM:pidotsigma}
\end{align}
where $I_2\left(m,\alpha,\beta,R \right)$ is defined in Appendix \ref{SM:AppIntegrals}. Summing up Eqs.~(\ref{SM:muanddelta}) and (\ref{SM:pidotsigma}), we get the final result for $U^m_{ij}$.

\section{Current conservation proof}\label{SM:AppCurrentConservationProof}

Below we prove that the expression for T-matrix in Eq.~(\ref{SM:Tmatrix}) is current-conserving. We must, therefore, prove that 
$$
T^\dag \cdot D \cdot T = D \quad\Leftrightarrow\quad \left( e^{-\frac{i}{\hbar} \tilde{U} D} \right)^\dag D e^{-\frac{i}{\hbar} \tilde{U} D} - D = 0.
$$
First, we note that
$$
\left( e^{-\frac{i}{\hbar} \tilde{U} D} \right)^\dag = e^{\left( -\frac{i}{ \hbar} \tilde{U} D \right)^\dag} = e^{\frac{i}{\hbar} D  \tilde{U} }.
$$
Using the definition of the matrix exponential we have
$$
e^{\frac{i}{\hbar} D \tilde{U} } D e^{-\frac{i}{\hbar} \tilde{U} D} - D = -D + \sum\limits_{k,n=0} \frac{1}{k!n!}\frac{i^k (-i)^n}{\hbar^{k+n}} (D\tilde{U})^k D(\tilde{U} D)^n = -D + \sum\limits_{m=0}^\infty \frac{i^m}{\hbar^m m!}\sum\limits_{k=0}^m (-1)^{m-k} C_m^k (D\tilde{U})^k D(\tilde{U} D)^{m-k}
$$
The inner sum for $m=0$ gives us $D$ which is cancelled by $-D$ outside of sums. It is easy to see that for $m=1$ and $m=2$ the inner sum equals $0$, therefore, we assume that the inner sum is always zero. We prove this by mathematical induction. We assume that the statement is correct for $m=N$, and we prove that it implies that the statement is also correct for $m=N+1$ using the relation for binomial coefficients, namely $C_{N+1}^k = C_{N}^k + C_{N}^{k-1}$.

\section{Conductance as a function of the chemical potential}\label{App:AnalyticalConductanceCurves}

\begin{figure}[h!]
	\begin{center}
	\includegraphics[width=0.49\columnwidth]{mudepdifffluxes.pdf}
	\includegraphics[width=0.49\columnwidth]{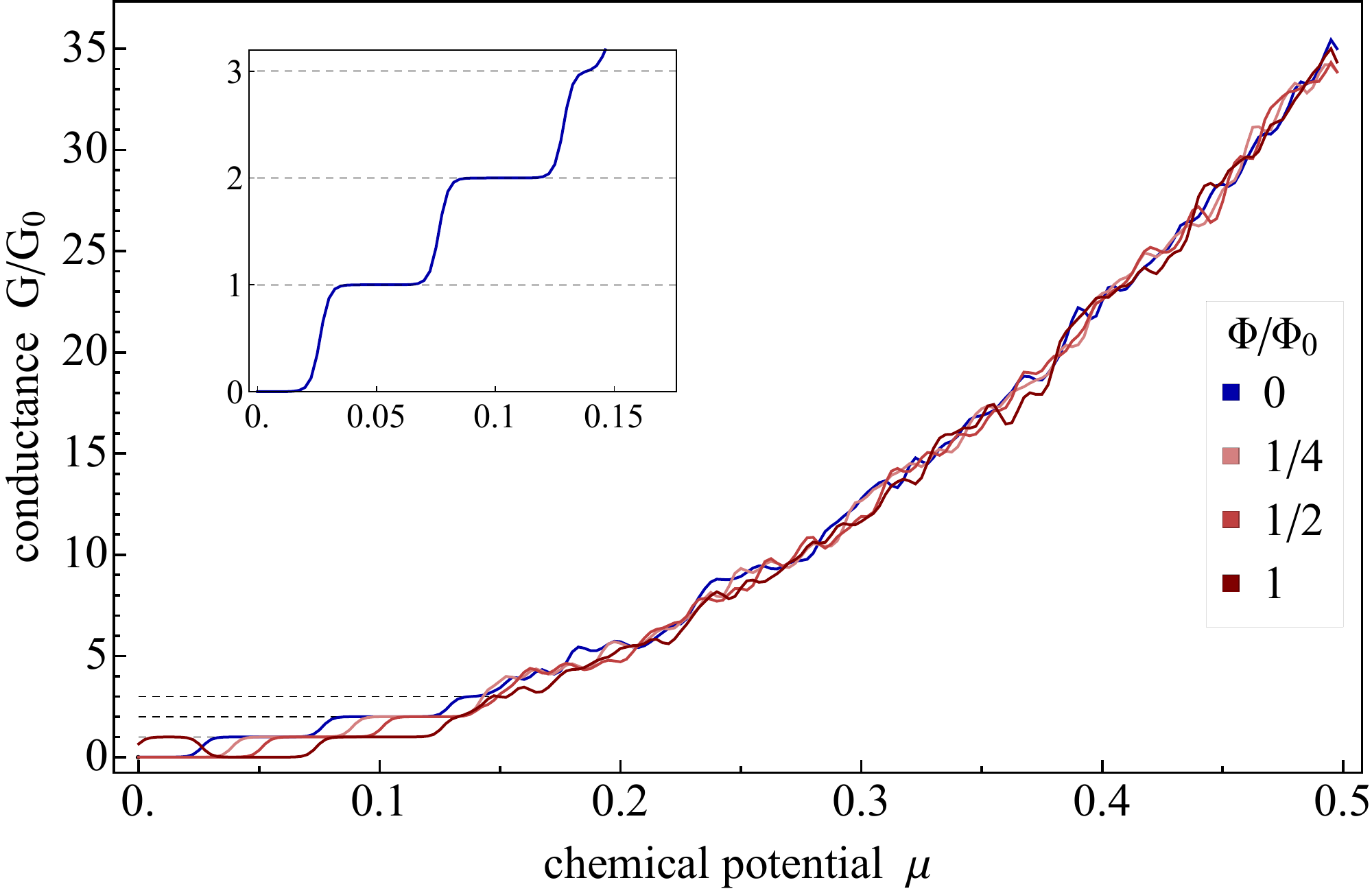}
	\caption{Conductance $G$ of a Weyl nanowire in the units of the conductance quantum $G_0$ plotted as a function of the chemical potential in the sample, at $T = 23\,$K. Different colors correspond to different values of the magnetic flux penetrating the wire. Left panel: numerical calculation with the Kwant package. Right panel: analytical calculation with the transfer matrix approach. In the numerical simulations values of conductance are most of the time lower due to the fact that there is nonzero scattering between the cones, whereas in the analytical model the cones are disconnected. In order to recover the correct transport behavior in the analytical model, we set the conductance to zero by hand at those values of the chemical potential that do not cross any bands in the sample. Note that such a treatment is necessary, since there is no scattering between the cones intrinsically woven into the model.	
	}
	\label{SM:figcondmudepdifffluxes}
	\end{center}
\end{figure}

To demonstrate that the analytical approach to the problem yields qualitatively the same results as the numerical simulations with the Kwant package, we present in Fig.~\ref{SM:figcondmudepdifffluxes} the conductance curves at different values of the magnetic flux, penetrating the wire. Exactly as in the main text, here we add thermal broadening  corresponding to $T = 23\,$K. Since the electron charge $e$ and the reduced Planck constant $\hbar$ were set to unity in the analytical calculation, one quantum of flux $\Phi_0 \equiv h/e = 2\pi \hbar/e = 2\pi$ corresponds to having $B = \Phi_0/\pi R^2 = 2/R^2 = 2/20^2 = 1/200$.

\end{document}